\documentclass[12pt]{article}
\usepackage{graphicx}
\usepackage{bm}
\usepackage{amsmath}
\usepackage{times}
\usepackage{color}
\usepackage{lineno}

\usepackage[superscript,nomove]{cite}
\usepackage{setspace}
\usepackage{caption}
\usepackage{titlesec}
\titleformat{\section}{\normalsize\bfseries}{\Alph{section}.}{0em}{}
\newenvironment{sciabstract}{%
\bf}{}

\let\oldequation\equation
\let\oldendequation\endequation
\renewenvironment{equation}
{\linenomathNonumbers\oldequation}
{\oldendequation\endlinenomath}

\title{Giant Self Spin-Valve Effect in the Kagome Helimagnet}

\author
{Xitong Xu$^{1,2\ast\dag}$, Yonglai Liu$^{1,2\ast}$, Kesen Zhao$^{1,2\ast}$, Che-Min Lin$^{3\ast}$,\\ Miao He$^{1,2}$, Haitian Zhao$^{1,2}$, Qingqi Zeng$^{1}$, Yubin Hou$^{1}$,\\ Qingyou Lu$^{1}$, Ding-Fu Shao$^{4}$, Shuang Jia$^{5,6,7}$, Haifeng Du$^{1}$,\\ Wenjie Meng$^{1\dag}$, Tay-Rong Chang$^{3,8,9\dag}$ \& Zhe Qu$^{1,2\dag}$\\
\\
\normalsize{$^{1}$Anhui Key Laboratory of Low-Energy Quantum Materials and Devices,}\\
\normalsize{High Magnetic Field Laboratory, HFIPS,}\\
\normalsize{Chinese Academy of Sciences, Hefei, Anhui 230031, China}\\
\normalsize{$^{2}$Science Island Branch of Graduate School,}\\
\normalsize{University of Science and Technology of China, Hefei, Anhui 230026, China}\\
\normalsize{$^{3}$Department of Physics, National Cheng Kung University, Tainan 701, Taiwan}\\
\normalsize{$^{4}$Key Laboratory of Materials Physics, Institute of Solid State Physics,}\\
\normalsize{HFIPS, Chinese Academy of Sciences, Hefei 230031, China}\\
\normalsize{$^{5}$International Center for Quantum Materials, School of Physics,}\\
\normalsize{Peking University, Beijing 100871, China}\\
\normalsize{$^{6}$Interdisciplinary Institute of Light-Element Quantum Materials and Research Center }\\
\normalsize{for Light-Element Advanced Materials, Peking University, Beijing 100871, China}\\
\normalsize{$^{7}$CAS Center for Excellence in Topological Quantum Computation,}\\
\normalsize{University of Chinese Academy of Sciences, Beijing 100190, China}\\
\normalsize{$^{8}$Center for Quantum Frontiers of Research and Technology (QFort), Tainan 701, Taiwan}\\
\normalsize{$^{9}$Physics Division, National Center for Theoretical Sciences, Taipei 10617, Taiwan}\vspace{0.4cm}\\
\normalsize{$^{\ast}$These authors contributed equally: Xitong Xu, Yonglai Liu, Kesen Zhao and Che-Min Lin}\vspace{0.2cm}\\
\normalsize{$^{\dag}$E-mail: xuxitong@hmfl.ac.cn; wjmeng@hmfl.ac.cn;}\\
\normalsize{u32trc00@phys.ncku.edu.tw; zhequ@hmfl.ac.cn}
}

\begin{document}
\baselineskip26pt
\captionsetup[figure]{labelfont={bf},name={Fig.},labelsep=period}
\maketitle
\clearpage

\begin{sciabstract}
%\linenumbers
{}{Kagome magnets can combine non-trivial band topology and electron correlations}, offering a versatile playground for various quantum phenomena.
In this work we propose that kagome magnets with frustrated interlayer interactions can intrinsically support a self spin-valve effect, and {}{experimentally confirm this in} the kagome helimagnet TmMn$_6$Sn$_6$.
Under a magnetic field perpendicular to the helical axis, {}{using magnetic force microscopy we observed} stripe domains that stack strictly along the helical axis, which {}{we attribute to} the stability loss of the kagome helimagnetic state.
Such a domain pattern spontaneously mimics the artificial multilayered structure in traditional spin valves, which, combined with the high spin polarization, leads to a giant magnetoresistance (GMR) ratio over 160\%.
This discovery opens an avenue to realize inherent spin valves in a variety of quantum magnets, and can hold promise in future spintronics.
\end{sciabstract}
\bigskip
%\section*{Introduction}
%\linenumbers

The kagome lattice, comprised of corner-sharing triangles as shown in Fig.~\ref{f1}(a), has been intensely studied in the search for emergent quantum phenomena for over 70 years~\cite{10.1143/ptp/6.3.306}.
The frustrated two-dimensional lattice hosts Dirac dispersion, van Hove singularities and flat band in the reciprocal space, which promotes a nontrivial band topology and strong electron correlations~\cite{Yin2021,Yin2022Topological,Xu2023,Wang2023}.
Depending on the band filling, the spin-orbit coupling strength and magnetism, this model system can sustain exotic quantum phases including the Chern insulator, the fractional quantum Hall state, electronic or magnetic instabilities, and unconventional superconductivity, etc~\cite{PhysRevLett.61.2015,PhysRevLett.106.236802,PhysRevLett.106.236804,PhysRevLett.35.120,PhysRevLett.69.1608,PhysRevLett.84.143,PhysRevLett.87.187004}.
In addition to the hopping within the kagome plane, the interaction along the stacking direction of kagome layers in real materials exerts further manipulation on the electrons.
Here we demonstrate that frustrated interlayer exchange couplings in kagome magnets can induce a spin-dependent transport phenomenon of application potentials, the self spin-valve effect (Fig.~\ref{f1}(b)).

\section*{Self spin valve in the kagome helimagnet.}
\begin{figure*}[htbp]
    \begin{center}
        \includegraphics[clip, width=0.75\textwidth]{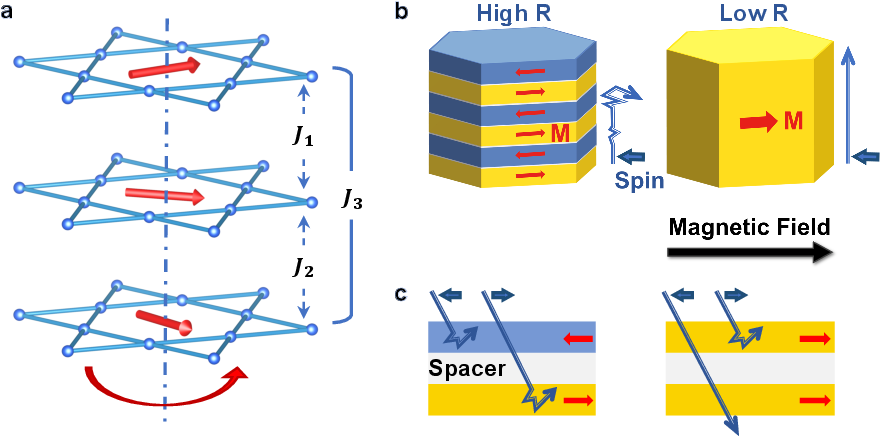}\\[0pt]  % insert figure
    \caption{{\bf Realizing self spin valve in the kagome helimagnet.}
{\bf a,} The kagome lattice and frustrated interlayer exchange couplings in kagome helimagnets.
{\bf b,} Proposed self spin valve in crystalline bulk kagome materials, and comparison with
{\bf c,} Traditional giant magnetoresistance (GMR) in ferromagnet/nonmagnet/ferromagnet trilayers.}
        \label{f1}
    \end{center}
\end{figure*}

The spin valve, widely used in magnetic sensors and hard-disk read-heads, utilizes a spin-dependent transport scenario called the giant magnetoresistance (GMR)~\cite{DIENY1994335,Thompson_2008,TSYMBAL2001113}.
Its functional structure involves alternating thin films of ferromagnet/nonmagnetic spacer/ferromagnet as shown in Fig.~\ref{f1}(c), where the relative spin orientation of the two magnetic layers controls the scattering process and leads to the GMR.
This, however, relies on nanoscale deposition and patterning of artificially tailored thin-film heterostructures, or mechanically stacking of van der Waals heterojunctions with precise layer counting~\cite{TSYMBAL2001113,Sierra2021,https://doi.org/10.1002/adma.201903800}.
There have been efforts to reproduce the spin-valve effect in granular magnetic solids and bulk ruthenates~\cite{10.1063/1.353765,10.1063/1.353766,PhysRevLett.100.016604}, yet careful control of chemical distribution or composition is still needed.
For instance, optimal Cr-doping is needed to form one Cr-O layer every 2 or 3 Ru-O bilayers in Ca$_3$(Ru$_{0.83}$Cr$_{0.17}$)$_2$O$_7$, and the resultant GMR size is only 50\% ~\cite{PhysRevLett.100.016604}.

Kagome magnet turns out to be an answer to whether there exists a bulk spin valve with high performance and robustness.
We firstly consider the situation where neighbouring magnetic kagome layers, with easy-plane anisotropy, are crystallographically equivalent as illustrated in Fig.~\ref{f1}(a).
If the exchange integral $J_3$ for the next-nearest-neighbor layers has opposite sign with respect to the nearest-neighbor exchange integral $J_1$ ($J_1=J_2$ in this case), and the condition $|J_3|>1/4|J_1|$ holds, the magnetic energy has a minimum for a spiral structure, with pitch angle between magnetic moments of nearest-neighbor layers being $\phi=\arccos\left(-J_1/4J_3\right)$.
This is the Yoshimori-type helimagnet stabilized by frustrated interlayer interactions~\cite{PhysRev.116.888,1959807}.
If the magnetic kagome layers are separated by different building blocks ($J_1\neq J_2$), there can exist multiple sets of pitch angles, leading to the appearance of double or even triple spiral structure~\cite{ROSENFELD20081898}.
Experimentally, helical states with short periods (typically $\lambda<10$~nm) have been observed in a wide range of Mn-based kagome magnets, including RMn$_6$Sn$_6$ (R = Er, Tm, Sc, Y, Lu), RMn$_6$Ge$_6$ (R = Tb, Dy), and their Ga-substitutes, etc~\cite{VENTURINI199135,venturini1993magnetic,VENTURINI1996102,LEFEVRE200284,ROSENFELD20081898,PhysRevMaterials.7.024404}.

In the chiral helimagnet, with Cr$_{1/3}$NbS$_2$ being a representative, the helimagnetic structure is protected by crystal chirality~\cite{PhysRevLett.108.107202}.
Under a magnetic field perpendicular to the helical axis, the ground state continuously evolves into a chiral soliton lattice~\cite{PhysRevLett.108.107202}.
In the Yoshimori-type centrosymmetric kagome helimagnet, however, chiral symmetry is preserved at the level of Hamiltonian in the absence of the Dzyaloshinskii-Moriya interaction.
The helical spin structure here, therefore, has no symmetry protectorate and is easily fragmented into multidomains under external stimuli~\cite{KISHINE20151}.
Especially, this instability, from competition between chiral magnetism and achiral Hamiltonian,  leads to a discontinuous transition from a helimagnetic structure to a fan or forced-ferromagnetic structure under magnetic field perpendicular to the helical axis~\cite{KISHINE20151}.
This can be seen from our numeric simulation of the magnetization curve for kagome magnets with the double-spiral ground state as shown in Supplementary Information (SI) Fig.~S1.
At some critical in-plane magnetic field, the rotation of one of the moments initiates an avalanche-like reorientation of the rest moments to reach the global energy minimum, which leads to a magnetic-history-dependent hysteresis between the field-sweep branches.
An important consequence is that the field polarized states can be sustained during field-sweep branches in the form of alternating domains as illustrated in Fig.~\ref{f1}(b).
The upper bound of the domain wall thickness is also limited by $\lambda$, avoiding the damping of GMR size due to spin precession across the domain walls~\cite{PhysRevLett.77.1580,Thompson_2008}.
These features in kagome helimagnets are the key ingredients to realize a large, bulk spin-valve effect without the necessity of spacing layers;
we thus dub it as a self spin-valve effect.

\section*{Material candidate.}
\begin{figure*}[htbp]
    \begin{center}
        \includegraphics[clip, width=1\textwidth]{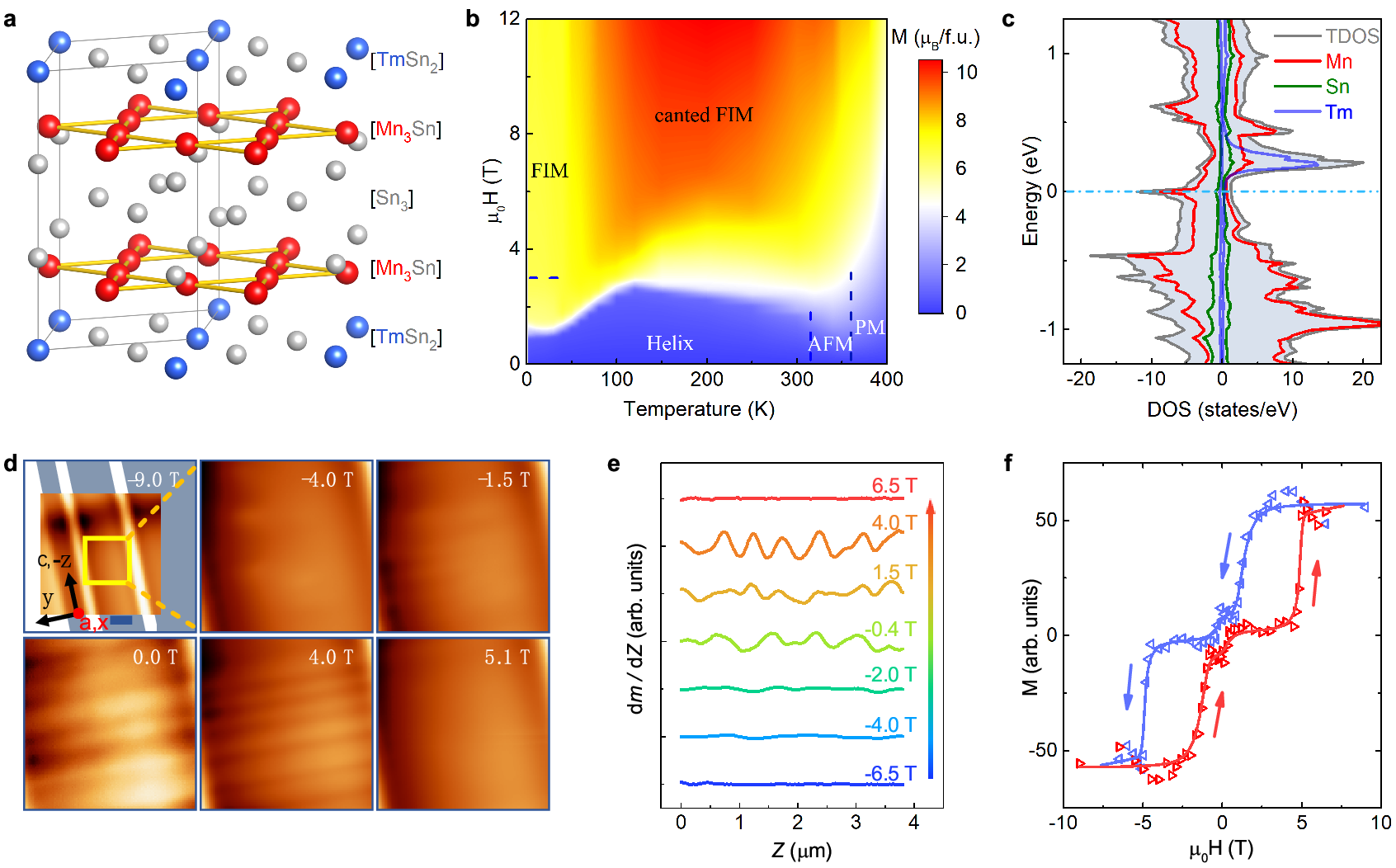}\\[0pt]  % insert figure
    \caption{{\bf Fragmented multidomains in TmMn$_6$Sn$_6$.}
{\bf a,} Crystal structure of TmMn$_6$Sn$_6$, which is composed of alternating slabs of [TmSn$_2$], [Mn$_3$Sn], [Sn$_3$], and [Mn$_3$Sn].
{\bf b,} Magnetic phase diagram of TmMn$_6$Sn$_6$ under an in-plane field, showing multiple phase areas including the low-field helical-ordered (Helix), antiferromagnetic (AFM) and paramagnetic (PM) regions, the high-field ferrimagnetic (FIM) and possible canted-FIM regions.
{\bf c,} Calculated density of states (DOS) in the FIM state.
{\bf d,} Magnetic force microscope (MFM) images of the spontaneous domain formation in TmMn$_6$Sn$_6$ at 5~K.
Magnetic field is swept from -9~T to {5.1~T} along the $x$ (crystalline $a$) direction (perpendicular to the paper plane).
The scale bar is 2~$\mu$m.
In the first panel, three samples with length around 20~$\mu$m (along the $z$ direction, or $c$ axis), thickness around 1~$\mu$m, and different widths around 1.5, 3, 6~$\mu$m (left to right) in series are shown.
The remaining 5 panels show local zoom-in.
{\bf e,} Derivative of the MFM strength along the $z$ (crystalline $c$) direction at representative fields.
{\bf f,} Magnetization profiles from integration of the local MFM strength.}
        \label{f2}
    \end{center}
\end{figure*}

We demonstrate TmMn$_6$Sn$_6$ to be one such material candidate for self spin valves.
As shown in Fig.~\ref{f2}(a), it crystallizes in the centrosymmetric, hexagonal HfFe$_6$Ge$_6$-type structure consisting of kagome slabs [Mn$_3$Sn] separated by two inequivalent [Sn$_3$] and [TmSn$_2$] slabs.
The interlayer exchange parameters between [Mn$_3$Sn] slabs fall into the Yoshimori-type frustration~\cite{riberolles2023new}, and TmMn$_6$Sn$_6$ behaves similarly to the non-magnetic R variants like YMn$_6$Sn$_6$ due to the relatively weak coupling between Tm and Mn~\cite{LEFEVRE200284,ghimire2020competing,PhysRevB.106.125107}.
According to the neutron scattering results~\cite{LEFEVRE200284}, the Mn sublattice displays an incommensurate double spiral structure below $\sim$330~K as shown in the phase diagram in Fig.~\ref{f2}(b).
The Tm sublattice also becomes magnetically ordered and forms another spiral structure which tends to be antiferromagnetically coupled with neighboring Mn moments.
In this manner the Mn and Tm spirals are referred to as a triple-spiral structure, and can be roughly described together by a propagating vector of (0, 0, $\sim$0.16) below 50~K, which corresponds to a short helix period of 5.5~nm\ ~\cite{LEFEVRE200284}.
In the existence of an applied magnetic field in the $ab$ plane, the Mn moments can be polarized with Tm moment antiparallel-aligned, forming a ferrimagnetic (FIM) state.
Our density functional theory (DFT) calculations confirm a large spin polarization factor (81\%) in the FIM state as shown in Fig.~\ref{f2}(c), which is crucial for a spin-dependent process to dominate in transport~\cite{5389307,Marrows2005}.

In order to trace possible domain structures in TmMn$_6$Sn$_6$, we have adopted the magnetic force microscopy (MFM) method which is sensitive to the perpendicular stray field gradient of the detected surface (See Methods section).
Interestingly, we observe stripe domains in focused-ion-beam (FIB) milled TmMn$_6$Sn$_6$ thin plates under an in-plane magnetic field.
The domain pattern is apparently asymmetric with respect to field, appearing at around -1.5~T and persisting to 5~T during a sweep-up circle (Fig.~\ref{f2}(d)).
Note that the orientation of the stripe domains is insensitive to the sample geometry or size, and the domains always stack along the helical $c$ axis (SI Figs.~S3 and S4).
Typical domain width is on the order of 300~nm as shown in Fig.~\ref{f2}(e).
From the integration of the local MFM strength, we further obtained the relative magnetization profiles in Fig.~\ref{f2}(f), which is similar to our calculated hysteresis in SI Fig.~S1.
The high spin polarization and the fragmented multidomains in TmMn$_6$Sn$_6$, therefore, pave the way for the appearance of the self spin-valve effect.

\section*{GMR in TmMn$_6$Sn$_6$.}
\begin{figure*}[htbp]
    \begin{center}
        \includegraphics[clip, width=0.93\textwidth]{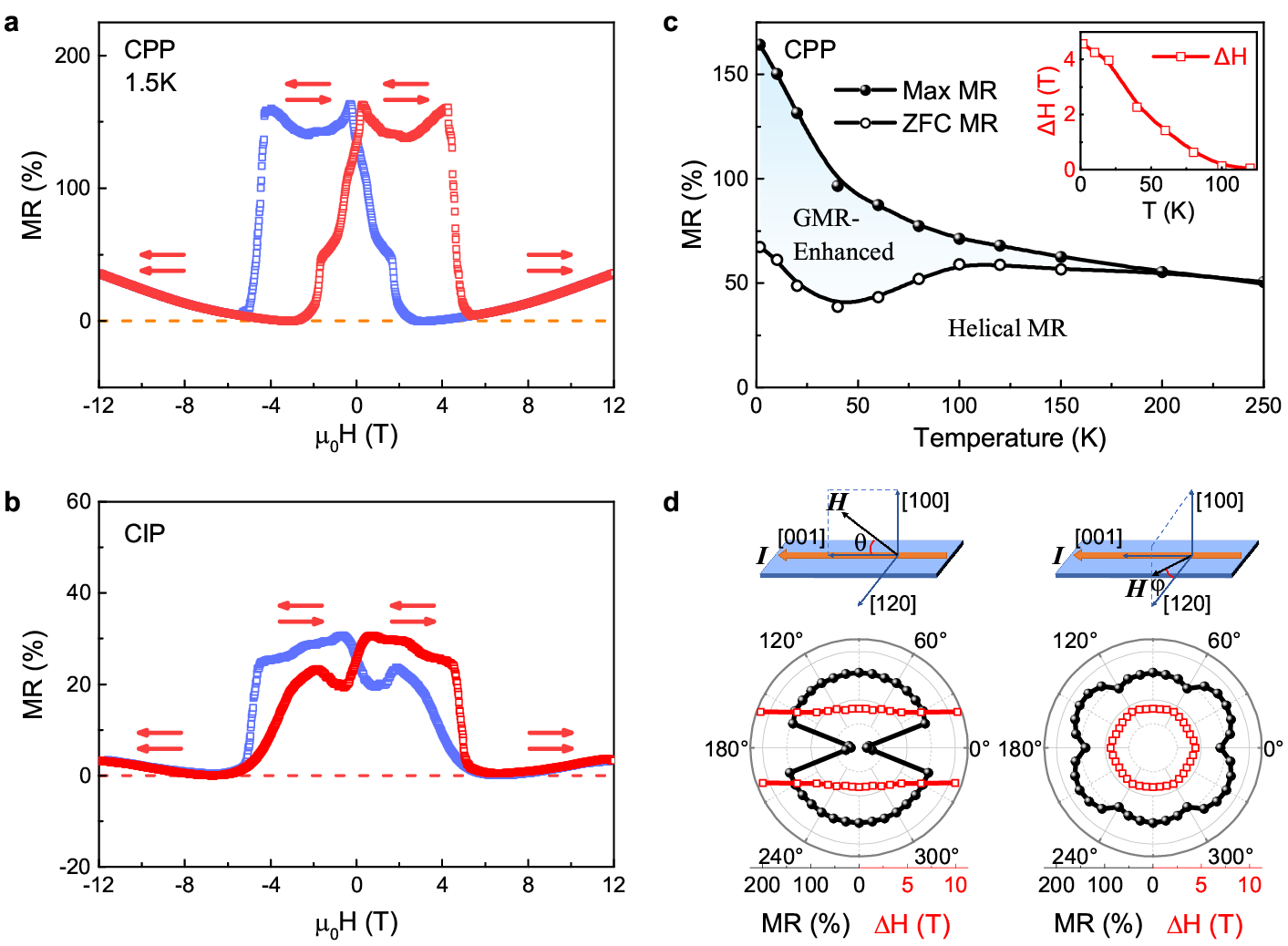}\\[0pt]  % insert figure
    \caption{{\bf Giant self spin-valve effect in TmMn$_6$Sn$_6$.}
{\bf a,} Self spin-valve effect for the current perpendicular to the domain plane (CPP) configuration at 1.5~K.
{\bf b,} Self spin-valve effect for the current in domain plane (CIP) configuration at 1.5~K.
{\bf c,} Temperature evolution of the magnetoresistance (MR) magnitudes.
Solid circles represent the maximal MR size whereas the open circles compare the unpolarized, zero-field-cooled (ZFC) resistivity to the minimal resistivity under external fields.
Inset shows temperature dependence of the hysteresis field $\Delta H$ between the sweep-up and down branches.
{\bf d,} Upper panels: Sample configurations for field $H$ rotating away from the current in the $ca$ plane and perpendicular to the current, respectively.
Lower panels: Corresponding angular dependence of GMR size and $\Delta H$.}
        \label{f3}
    \end{center}
\end{figure*}

We have measured the resistance of TmMn$_6$Sn$_6$ thin plates in different directions with field $H$ perpendicular to the current $I$ (SI Figs.~S5).
For $H\parallel c$, $I\parallel ab$ configuration, the change of resistance with respect to $H$ is small and positive.
For $H\perp c$, however, both the current perpendicular to domain plane ($I\parallel c$, CPP) and current in domain plane ($I\parallel ab$, CIP) designs exhibit a remarkable hysteresis as shown in Fig.~\ref{f3}(a) and (b), which is typical for a spin valve and is in good accord with the measured magnetization profiles.
In the region where the moments are polarized by the field, the MR ratio (defined as the relative change with respect to the minimal value in the polarized state) is small.
However, in the region where the domain structures appear, the resistance is greatly enhanced, and MR in the CPP setup is as large as 164\% at 2~K.
Magnetic-history-dependent memory effect, similar to traditional spin valves~\cite{Jedema2001}, also appears as shown in Fig.~S7 in SI.
Even in the CIP structure, the MR in TmMn$_6$Sn$_6$ device is still over 30\%.
We have compared the magnitude of GMR from different mechanisms in SI Fig.~S8.
The MR ratio in the CPP configuration is comparable to state-of-arts epitaxial GMR spin valves~\cite{DIENY1994335,10.1063/1.111253}, and much higher than those in granular Co$_{20}$Ag$_{80}$ and Cr-doped bulk Ca$_3$Ru$_2$O$_7$ ~\cite{10.1063/1.353765,10.1063/1.353766,PhysRevLett.100.016604}.
This value even competes with the tunneling magnetoresistance of many magnetic tunneling junctions~\cite{Sierra2021,https://doi.org/10.1002/adma.201903800,Elahi_2022}, yet maintaining a significant smaller resistance-area product, which is crucial for higher operating frequencies and lower Johnson and shot noises in practical devices~\cite{nagasaka2001cpp}.
{}{We note that the observed GMR with apparent memory effect and the underlying domain-based physics are entirely distinguished from the results in Ref.~\cite{PhysRevB.106.125107} and layered magnetic systems like the manganites~\cite{annurev.matsci.30.1.451}.}

As the temperature rises, the MR magnitudes monotonously decrease and the hysteresis field $\Delta H$ between the sweep-up and down branches vanishes above 100~K as shown in Fig.~\ref{f3}(c), probably due to enhanced scattering probabilities and thermal fluctuations.
We also compare the resistance in the multidomain states with respect to the one in the pure helical states obtained from a zero-field-cooled (ZFC) procedure in Fig.~\ref{f3}(c).
As expected, the multidomains generate a way larger GMR ratio than the helical states at low temperatures, suggesting that the alternating domain patterns are more efficient in tuning the spin-dependent transport.

The angle dependence of the self spin-valve effect is studied as well in Fig.~\ref{f3}(d).
For $H$ rotating in the $ca$ plane ($\theta=0^\circ$ when $H\parallel c$), $\Delta H$ shows a $1/\sin\theta$ behavior, suggesting a strong easy-$ab$-plane anisotropy.
The size of MR is almost unchanged at various angles except when the $ab$ plane component of $H$ is not enough to alter the domain orientations.
For $H$ rotating in the $ba$ plane, $\Delta H$ possesses a six-fold symmetry from the Mn kagome lattice due to the crystal-field effect~\cite{riberolles2023new}.
The maximal MR occurs when the field is along to the crystalline $b$ axis for our setup, amounting up to 175\%.

\section*{Discussion}

The above observations and analyses are not limited to the FIB-milled TmMn$_6$Sn$_6$ samples.
We have also observed similar domain patterns and corresponding self spin-valve effect in millimeter-sized crystals (SI Fig.~S12).
Nevertheless, the GMR ratio in the latter is smaller (88\% at 2~K) for the CPP configuration, and the critical field at which the domains are fully aligned decreases to 1.9~T.
We attribute the discrepancies to an over-enlarged domain width with respect to the spin diffusion length in millimeter-sized samples (SI Fig.~S12(a)) and the reduced pinnings from boundaries~\cite{Bass_2007,PhysRevB.48.10335}.
As the avalanche-like reorientation in Fig.~\ref{f2}(f) (also SI Fig.~S1) depends on detailed exchange parameters, the Ga-inclusion during FIB milling might also have subtle influences on the critical field~\cite{bachmann2020manipulating}.
We leave these for future explorations.

Though we have treated the Mn and Tm spirals as a whole throughout our investigations, the rare earth site matters due to its anisotropy and its exchange coupling with the Mn kagome layers, offering an easy tunability of the transition temperatures, the detailed magnetic structures and band structures in RMn$_6$Sn$_6$ or RMn$_6$Ge$_6$~\cite{VENTURINI199135,PhysRevLett.126.246602,lee2022interplay,PhysRevMaterials.7.024404}.
We have also considered the inclusion of Tm moment and its anisotropy in a triple-spiral model in the SI Note~3.
Nevertheless, the magnetization profiles are very similar to the double spiral model due to the relative weak Tm-Mn interaction.

For practical account, the critical field to switch the spin valve should be small.
Theoretically, this effect can be tuned to near-zero magnetic field by manipulating the magnetic interaction of kagome helimagnets in the parameter space as shown in Fig.~S15.
We have also taken the thermal effect into consideration in SI Note~2.
It is shown in Fig.~S16 that for the fixed exchange parameters, the overall effect of increasing temperature is to decrease the field needed to alter magnetic domains from parallel to antiparallel and antiparallel to parallel states.
These findings therefore suggest that the kagome helimagnets, in principle, offer low-field controllability at room temperature.

In short, we have considered the domain degrees of freedom and put forward a self spin-valve effect in a wide range of kagome magnets with frustrated interlayer interactions, with the R166 family being a representative.
Especially, we show that in TmMn$_6$Sn$_6$, the GMR ratio can surpass 160\%.
Considering the vast variety of material candidates and the great tunability from rare-earth engineering, our finding may assist the flourish of spintronic devices in quantum magnets.

\bigskip
\bigskip

\section*{Methods}
\paragraph*{Material growth and characterization}
Single crystals of TmMn$_6$Sn$_6$ were synthesized via a standard tin flux method~\cite{PhysRevLett.126.246602}.
Tm lumps, Mn pieces and Sn grains with a molar ratio of 1: 6: 20 were packed into an alumina crucible which was then sealed in a fused quartz tube under vacuum.
The mixture was heated to 1000$^\circ$C, cooled down to 600$^\circ$C over 3 days and then centrifuged to remove residual flux.
Magnetic measurements on single-crystalline samples were performed in the Quantum Design PPMS-14~T and MPMS-3.

\paragraph*{Device fabrication and transport measurement}
Thin plates of TmMn$_6$Sn$_6$ were cut out from single crystals using the FEI Helios NanoLab FIB and patterned into standard Hall bars, with typical size around $8\times1\times1\ \mathrm{\mu m^3}$.
Electrical contact was made by FIB-assisted platinum deposition onto a silicon stage with pre-patterned gold electrodes.
Typical SEM picture of the device is shown in SI Fig.~S4.
Electrical transport measurements were performed in the 12~T C-MAG Teslatron PT system with the SynkTek multichannel Lock-in.

\paragraph*{MFM measurement}
The MFM experiments were conducted using an in-house-built cryogenic magnetic force microscope using commercial piezoresistive cantilevers (Quantum Design model PRSA-L400-F30-Si-sdPCB) which have a resonance frequency of about 35~kHz and a $Q$ factor about 10$^4$ at 5~K).
One side of the tips was coated with 5~nm Ti and 40~nm Co films using electron beam evaporation.
The magnetic force microscope was controlled by a commercial RHK controller with a built-in phase-locked loop module.
The signal used to image was a shift of the cantilever's resonant frequency (dF), which is proportional to the out-of-plane stray field gradient.
During each scan, a topographic image was first obtained using a contact mode to compensate the sample surface tilting along the fast and slow-scan axes, after which the tip was lifted by 100~nm and the MFM images were collected in a constant height mode.
When the net magnetic moment of a domain lies parallel/antiparallel to the external field $H$, the magnetic force is attractive/repulsive, leading to a negative/positive dF (dark/light areas in MFM images).
MFM images were analyzed using the Gwyddion software.
MFM measurements were performed both on the FIB samples of different sizes and orientations (SI Figs.~S2 and S3) and highly polished (100) surface of millimeter-sized bulk samples (SI Fig.~S12).

\paragraph*{DFT calculation}
The bulk band structures of TmMn$_6$Sn$_6$ (Fig.~S11) were computed using the projector augmented wave method as implemented in the VASP package~\cite{P.E.1994,PhysRevB.59.1758,KRESSE199615} within the GGA+U scheme~\cite{PhysRevLett.77.3865}.
The experimental lattice parameters and the in-plane ferrimagnetic (FIM) magnetic configuration were used.
The spin-polarized calculation was included self-consistently in the calculations of electronic structures with a $\Gamma$-centered k-point mesh of $15\times15\times9$.
We employed the same methodology in Ref.~\cite{lee2022interplay}, utilizing the GGA+U approach with a U$_{\mathrm{Tm}-4f}$ value of 6.4~eV.
The computed spin magnetic moment for Tm is 1.917~$\mu_\mathrm B$, consistent with the Hund's rule prediction of 2~$\mu_\mathrm B$.
The average spin magnetic moment for Mn is determined to be 2.38~$\mu_\mathrm B$, which is in close agreement with the results reported in Ref.~\cite{lee2022interplay}.
{}{The visualization of models was performed by using the VESTA code~\cite{Momma:db5098}.}

\bigskip

\paragraph*{Magnetic hysteresis calculation}
A double-spiral model was adopted for the calculation of the magnetic hysteresis~\cite{ROSENFELD20081898}, where the pitch angle $\phi_n$ for moment in the $n$th Mn layer in the zero-field spiral state is

\vspace{-0.5cm}
\begin{equation}
\begin{aligned}
\phi_n=\left\{
 \begin{array}{ll}
 k\Phi,       & n=2k,\\
 k\Phi+\delta,& n=2k+1. \\
 \end{array}
\right.
\end{aligned}
\end{equation}
The pitch angle is modulated under an external in-plane magnetic field, and the total magnetic energy is

\vspace{-0.5cm}
\begin{equation}
\begin{aligned}
E=&-\frac{1}{N}\sum_{k=0}^{N-1}\{BM_s\left[\cos(\phi_{2k})+\cos(\phi_{2k+1})\right]+J_1M_s^2\cos(\phi_{2k+1}-\phi_{2k})\\
&+J_2M_s^2\cos(\phi_{2k+1}-\phi_{2k+2})+J_3M_s^2\left[\cos(\phi_{2k+2}-\phi_{2k})+\cos(\phi_{2k+3}-\phi_{2k+1})\right]\},
\end{aligned}
\end{equation}
where $M_s$ is the saturation magnetization, and the exchange parameters ($x=J_2/J_1=0.243$, $y=J_3/J_1=-0.12$) used here are similar to the reported values in Ref.~\cite{ROSENFELD20081898}.
The optimal magnetic moment angles during the field sweeps were solved numerically by minimizing the energy, and were subsequently converted into magnetization as shown in Fig.~S1.

We also derived the exchange coupling values directly from a Heisenberg model~\cite{PhysRevB.78.224517}.
The Heisenberg Hamiltonian is written in the following form:

\vspace{-0.5cm}
\begin{equation}
\begin{aligned}
H=J_1\sum_{ij}\overrightarrow{S_i}\cdot \overrightarrow{S_j}+J_2 \sum_{ij}\overrightarrow{S_i}\cdot \overrightarrow{S_j}+J_3 \sum_{ij}\overrightarrow{S_i}\cdot \overrightarrow{S_j},
\end{aligned}
\end{equation}
where $S_{i/j}$ is the spin of each Mn ion.
Considering two Mn layers associated with the exchange coupling $J_1$ (as illustrated in Fig.~1(a) in the main text), we have $\sum_{ij}\overrightarrow{S_i}\cdot \overrightarrow{S_j}=\pm3|S^2|$ for the parallel/antiparallel alignment.
Consequently, this can be expressed as $J_1=\left(E_{F,1}-E_{AF,1}\right)/6|S^2|$, where $E_{F,1}-E_{AF,1}$ is the energy difference between the parallel and antiparallel alignments for a pair of nearest neighboring moments.
$J_2$ and $J_3$ were calculated in the same methodology.
The exchange parameters were found to be $x = 0.2449$ and $y = -0.1203$ from the DFT calculation, in remarkable correspondence to the values used in our simulated hysteresis curves.
\bigskip

\section*{Data availability}
The data that support the findings of this study are included in the published article and its Supplementary Information files.
These data are also available from the corresponding authors upon request.

\bigskip

\paragraph*{\normalsize Acknowledgments}
We thanks Prof Jia Li at ICQM, PKU for instructive discussions.
This work was supported by the National Key R \& D Program of China grant number 2022YFA1403603,
%2023YFA1407300 and 2023YFA1607701,
National Natural Science Foundation of China grant numbers U2032213, 12104461, 12374129, 12141002, 12225401 and 12304156,
Chinese Academy of Sciences under contract numbers YSBR-084, JZHKYPT-2021-08 and XDB28000000.
This work was also supported by Anhui Provincial Major S \& T Project (s202305a12020005), Anhui Provincial Natural Science Foundation No. 2408085J025.
A portion of this work was supported by the High Magnetic Field Laboratory of Anhui Province under Contract No. AHHM-FX-2020-02.
T.-R.C. was supported by National Science and Technology Council (NSTC) in Taiwan (Program No. MOST111-2628-M-006-003-MY3 and NSTC113-2124-M-006-009-MY3), National Cheng Kung University (NCKU), Taiwan, and National Center for Theoretical Sciences, Taiwan.
This research was supported, in part, by the Higher Education Sprout Project, Ministry of Education to the Headquarters of University Advancement at NCKU.
T.-R.C. thanks the National Center for Highperformance Computing (NCHC) of National Applied Research Laboratories (NARLabs) in Taiwan for providing computational and storage resources.

\paragraph*{\normalsize Author contributions}
X.X. and Z.Q. conceived the project;
X.X. and Y.L. fabricated the devices and conducted the transport experiments in consultation with D.S., S.J., H.D., W.M. and Z.Q.;
K.Z. {}{and Y. H.} conducted the MFM measurements in consultation with X.X., Q.L., and W.M.;
Y.L., M.H., H.Z. and Q.Z. synthesized and characterized the bulk samples;
C.-M..L. carried out the theoretical analysis in consultation with T.-R.C.;
X.X. and Y.L. performed the data analysis and figure development;
X.X. wrote the paper with contributions from all authors;
Z.Q. supervised the project.
All authors discussed the results, interpretation and conclusion.

\paragraph*{\normalsize Competing interests}
The authors declare no competing interests.

\clearpage

\nolinenumbers
\section*{Figures}
\clearpage

\clearpage

\clearpage

\end{document}